# Ergonomics Integrated Design Methodology using Parameter Optimization, Computer-Aided Design, and Digital Human Modelling: A Case Study of a Cleaning Equipment

Running Title: Ergonomic Design Optimization Methodology

Neelesh Kr. Sharma[a]* Mayank Tiwari[a], Atul Thakur[a], and Anindya K. Ganguli[b]

[a]Department of Mechanical Engineering, Indian Institute of Technology, Patna, Bihar, 801106, India;

[b]Faculty in Ergonomics (Part-time), PG Section, Ram Mohan College, University of Calcutta, Kolkata, West Bengal, 700073, India.

***CORRESPONDING AUTHOR:** Neelesh Kr. Sharma,

Department of Mechanical Engineering, Indian Institute of Technology, Patna, Bihar, 801106, India. (phone: +91-7982211363; e-mail: 1821me12@iitp.ac.in).

WORD COUNT (Excluding references and abstract): 5434

ABSTRACT WORD COUNT: 161

NUMBER OF FIGURES: 11

NUMBER OF TABLES: 3

# Ergonomics Integrated Design Methodology using Parameter Optimization, Computer-Aided Design, and Digital Human Modelling: A Case Study of a Cleaning Equipment

**Abstract**

Challenges of enhancing productivity by amplifying efficiency and man-machine compatibility of equipment can be achieved by adopting advanced technologies. This study aims to present and exemplify methodology for incorporating ergonomics pro-actively into the design using computer-aided design and digital human modeling-based analysis. The cleaning equipment is parametrized to detect the critical variables. The relations are then constrained through the 3DSSPP software-based biomechanical and experimental analysis using a prototype. MATLAB and Minitab software is used for optimizing efficiency while satisfying the established constraints. The experiment showed nearly 67%, 120%, and 241% successive improvement in the mechanical advantage in comparison to their immediate predecessors. A significant (6 point) reduction in rapid entire body assessment score has been observed in the final posture while working with the manipulator. 3DSSPP suggested that the joint forces during the actuation of the manipulator were acceptable to 99% of the working population. The study demonstrated the potential of the methodology in revamping the equipment for improved ergonomic design.

# 1 Introduction

The computer has revolutionized the product design and development process over the past two decades. The use of computers at the designing phase helps to save cost and money, as well as, reduces the direct human involvement improving its safety. The simulation of the equipment, Digital Human Modelling (DHM), and their interactions can help to enhance its efficiency, comfort, and competitiveness. Body posture, reach envelope, the field of view, clearances, loading on lumbar spine segments, and other anthropometry and biomechanics visualizations are provided by DHM and serve as basic inputs for design analysis and engineering decision making. Although the requirement and advantages are numerous, a methodology to channelize computer integration is still under development.

The methodologies that include computer-based optimization often lack the basic guidelines for selection and parametrization of the variable for it. A more generalized methodology lacks the element of optimization and proposes a perceptive technique for evaluation. Both these methodologies fail to provide a robust guide that includes the depth of the former with the flexibility of the latter.

Jeang et al. [1] discuss the technique for determining the optimum parameter. In the study, the computer-based modeling and optimization methodology was exemplified for bike-frame design. Tarallo et al. [2] presented an interactive design method for improving workplace health and safety. The effects of anthropometric variability at the workplace were discussed in detail. The participation of the end-user in design optimization could enhance the conceptualization of design [3].

Kisaalita et al. [4] presented a guide for developing user-friendly equipment. The human-centered design approach was utilized with the local population. Bhattacharjya and Kakoty [5] gathered the anthropometric data and analyzed its variation in the same locality. The data was then used to improvise the traditional equipment. These techniques are helpful to achieve user requirements and design evolution but lack the optimization of human well-being and efficiency.

A deeper insight suggests that human well-being and efficiency also suffer from trade-offs. Few researchers suggest that increasing well-being enhances efficiency [6,7], while others perceive a different vision towards it [8,9]. Such trade-offs should be tackled judiciously as an inclination towards one could affect the other severely. Compromising with human comfort may seem to be boosting productivity but in the long run, it eventually ends up costing more [10]. So, this study proposes a new methodology for ergonomic design optimization and exemplifies it for manual cleaning equipment.

Manual material handling (MMH) work is the main contributor to the annual musculoskeletal disorder (MSD) cases in the US, costing around 54 billion dollars due to lost injury days [11]. MMH operations are tasks that include carrying, lifting, lowering, pulling, and pushing objects. Carrying and lifting are among the most demanding physical task repeatedly performed by the cleaning workers [12]. A strong association has been documented between MMH work (repetitive tasks, tasks with huge force requirements, and tasks with awkward posture) and musculoskeletal disorders (MSDs) [13,14]. Low back pain (LBP) is the MSD most commonly resulting in disability and the leading cause for years lived with disability [15]. The waste management sector has a potentially high risk of LBP due to the work's extreme physical demand [16].

Power hoists, pallet converters, drum rotators, and lifting trucks significantly reduce manual effort but do not apply to all workplaces [17]. The motorized mechanical equipment may impose a significant cost for purchase and maintenance [18]. Besides, they may require explicit training and proficient workers for the operation. Small scale industries and low-income countries rely upon manually operated tools [19]. Ergonomically designed equipment like cylinder trolleys [20], guided vehicles [21], manual carts [22], bucket lifters [23] are found to be significantly improving the working condition of the operator during MMH tasks. Thus, the non-motorized mechanical equipment becomes a suitable substitute owing to their tendency to be generally smaller and cheaper.

The current study is initiated to present a robust methodology for the integration of ergonomics in design. The study also emphasizes the role of computer-aided design (CAD) and DHM based analysis in the development of manually operated equipment. The methodology here guides the iterative development of the manual cleaning equipment.

Section 1 conversed the necessity of the methodology. Section 2 presents the outline of the methodology, while section 2.3 discusses the optimization step of the methodology in detail. Section 3 discourses the thorough evaluation of the developed equipment along with its comparison with the priorly developed versions.

## 2  Methodology

2.1 Proposed methodology

The proposed methodology's goal is to determine the ergonomic equipment design by parameter optimization. The eight-step process, with a feedback loop helping to continuously improve the design, is shown in Figure 1. The methodology begins with (a) Survey- It is conducted to get an insight into the methodology, process, and equipment used while performing the job. It requires direct interactions with the end-user/worker, as the interviews with the supervisors often include systemic biases. In the absence of the survey data, previous studies could also be utilized for the insight. (b) Identify- It helps to filter out the ergonomic tasks to concentrate on the uncomfortable ones. It often includes exhaustive listing and quantification for further actions. (c) User data- Anthropometry and the strength of the end user are the most prominent factors for ergonomic intervention. This step directly affects the suitability and acceptability of the improved version. Alternately, nationwide surveys or book references could also be used for data collection. (d) Equipment data- This includes the specification of the existing equipment as well as information of similar devices researched/used in other locations, fields, and countries. (e) Brainstorming- A group of experts, designers, and workers gather to discuss the issues identified earlier for probable solutions consistent with the requirement. (f) Optimize- All the solutions from the session are modeled and parametrized to classify independent and dependent variables. The step could be performed experimentally using prototypes or virtually through simulations. (g) Evaluation- The latest version is assessed for efficiency and human well-being-related improvements. (h) Launch- This involves the preparation of adequate strategies, involving local workers and distributors, for instigating the intervention. This step could be skipped if commercial viability is not targeted.

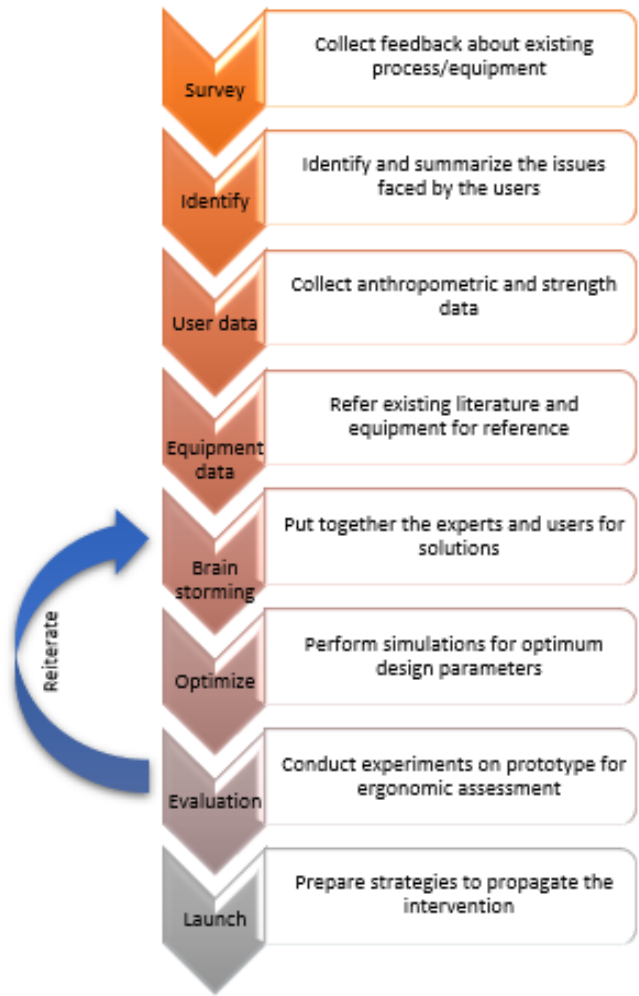

*Figure 1 The eight-step flowchart of the optimization methodology*

2.2 Cleaning equipment

The ergonomic intervention of the cleaning equipment is chosen in this study for demonstrating the eight steps of the methodology. (i) In the present study, out of the total 125 cleaning workers who were contacted, 92 participated in the feedback related to the existing cleaning practices [24]. (ii) The scraping and collection tasks during cleaning are identified as the critical activities, with back compression forces in the order of 2405 N and 3412 N respectively [25]. Dedicated equipment is warranted which could reduce the risks associated with such manual material handling tasks. The major requirements of the equipment as identified are listed below:

- The equipment should be operable by a single Indian worker (anthropometric requirement)
- The equipment should have the capacity and capability of the commercial equipment (feasibility requirement)
- The equipment should be suitable for Indian road and waste (usability requirement)

The criteria applied during the process of design selections are

- The complexity of design is minimized. Mechanical gears and electrical power are likely to increase the complexity.
- The design does not introduce any additional risk of MSDs
- The overall cost of the equipment is reduced. Expensive components or tools may eventually increase the final price.

(iii) The anthropometric data, like height and weight, is measured using a stadiometer (PrimeSurgicals, India) and weighing scale (HealthSense, India) with full-scale reading as 200±0.1 cm and 180±0.1 kg respectively. (iv) A review of fifty-nine manually operated equipment highlighted the scarcity of research in the ergonomics integrated designs within the countries in the low-income group [19]. The five major operations are expected to be performed by the selected cleaning equipment namely, a) picking the waste- the serial mechanism is held static using the locking link, and the mobile platform aid the collection of waste, b) lifting the waste- the mobile platform is held static using the stopper in wheels, and serial mechanism lifts the end effector, c) releasing the waste in the bin- tilt assures the smooth flow of material from bucket to the container, d) transporting to the desired location- rear castor wheels assist the easy maneuvering of the manipulator, and e) dumping- sliding side door and gradient in the base of container promotes the quick removal to form up to one-foot high material deposits (Figure 2). (v) The patented [26] version of the existing cleaning equipment is capable of picking up the material from the ground surface level (Figure 3). Version 2 introduced the sliders along with leg operation for lifting the bucket. Version 3 has been evolved considering user anthropometry with tiered handle arrangement. Details of the issues and their respective solutions obtained from the discussions for the cleaning equipment are summarized in Table 1. (vi) The process of optimization practiced while designing the equipment is discussed in detail in Section 2.3 ahead. (vii) The equipment has been successively improved and tested through experiments and three-dimensional static strength prediction program (3DSSPP) simulations [27]. The user data, collected earlier is used to generate population for the simulations while the $5^{th}, 50^{th,}$ and $95^{th}$ percentile member's anthropometry have been used to evaluate the man-machine interactions. The posture-based assessment tool, rapid entire body assessment (REBA), is simultaneously used for evaluation [28]. This step will be discussed in detail under the results heading. (viii) Launch- The local municipal bodies and cleaning staff were contacted for the deployment of the equipment. Eight pieces of equipment are developed in the preliminary phase for observing the response.

2.3 Optimization

The two major issues identified through the survey of cleaning activity are the collection and scarping tasks. The manual collection and scraping tasks are modified by the equipment as lifting and maneuvering. The lifting task is enhanced by optimizing mechanical advantage (MA) (Section 2.3.1) and the stroke length of the lifting handle (*Section 2.3.2.1*). The maneuvering task is improved by optimizing the maneuvering handle height (*Section 2.3.2.2*) and wheel size (Section 2.3.4).

2.3.1   Lifting mechanism

*2.3.1.1 Objective function*

The principal function of the equipment is lifting the material so it should be most efficient and ergonomic. The efficiency of a machine system can be measured in terms of MA which is the ratio of the load lifted to the effort applied. In Figure 4, AB is the connecting link inclined at an angle α with the horizontal, CD is the bucket link inclined at an angle γ with the vertical, and β is the angle between AB and CD. $F_I$ is the effort applied by the user, f is the force applied by link AB on link CD and W is the weight of bucket material. Using the moment balance equation on both the links AB and CD, we have Eq 1-2.

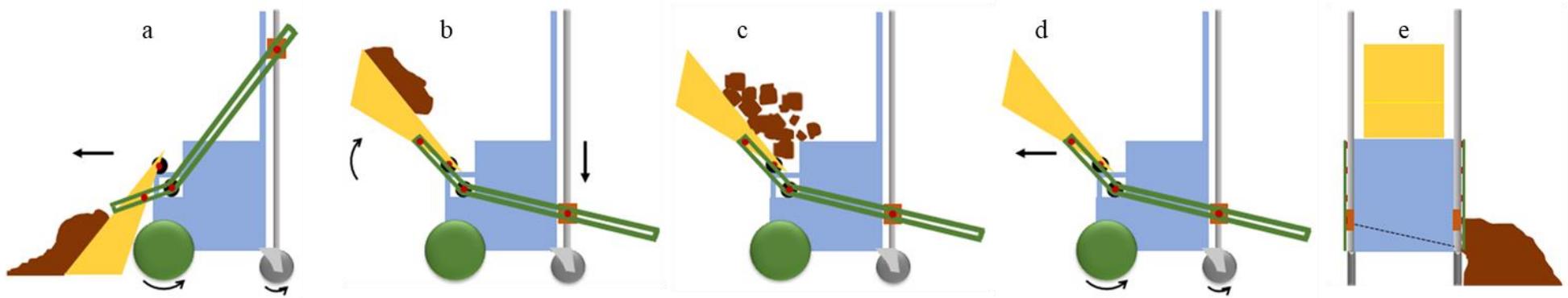

*Figure 2 Overview of the five cleaning operations: a) pick, b) lift, c) transfer, d) transport, and e) dump*

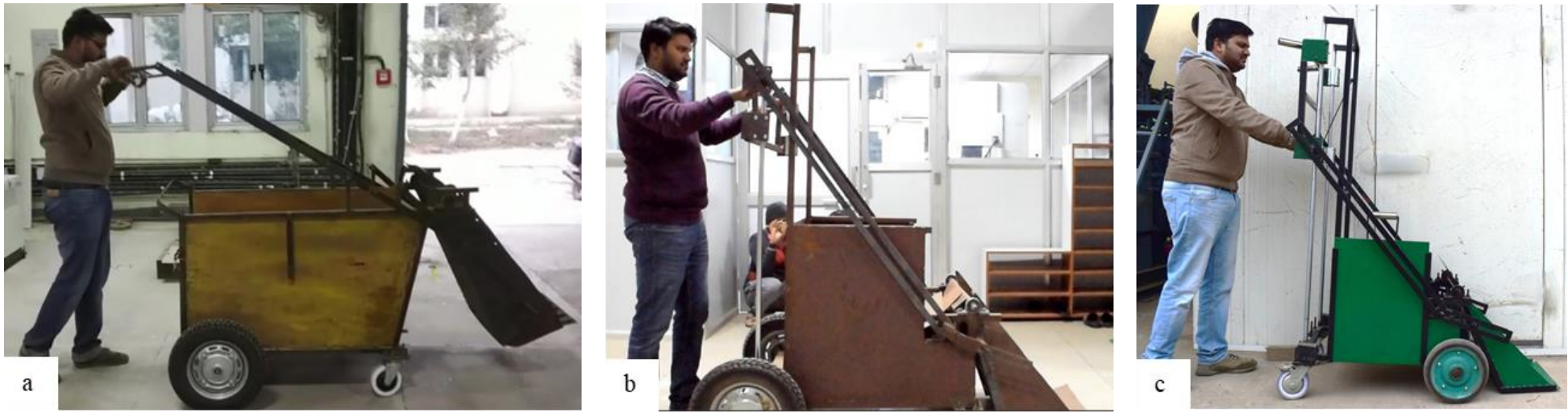

*Figure 3 Previous versions of equipment a) Initial Prototype, b) Modified Prototype, and c) Anthropometric Prototype*

*Table 1 Summary of the issues raised and solutions adapted for the design of equipment*

| S No | Feedback | Actions Discussed | Comments / Observations |
|---|---|---|---|
| | | Prototype Version 1 | |
| 1 | Difficulty in the bucket lift | Improve the mechanism and mechanical advantage | Is there any difference in the comfortable loading between genders |
| 2 | Handle height can be improved | Conduct anthropometric study and determine a suitable height | Workers prefer near erect posture of the operation |
| 3 | Try to reduce the push/pull effort | Improve tire shape and type | Which type of tire suits best? |
| 4 | Reduce the bending required for the manipulator | Conduct anthropometric study and determine a suitable stroke length | Can the stroke be made foot operable? |
| | | Prototype Version 2 | |
| 1 | Difficulty in turning | Improve tire arrangement | What should be the position of the swivel wheel for the pushing task? |
| 2 | Imbalance during foot operation | Switch to hands only operation | Can the stroke be divided into two repeated levels |
| 3 | Difficult to assemble/disassemble | Use cheaper parts that are easily available | It will also reduce the overall equipment cost |
| 4 | No dumping system | Include an easily operable waste removal system | Cost-effective, contact-free method |

| S No | Feedback | Actions Discussed | Comments / Observations |
| --- | --- | --- | --- |
| 5 | Difficulty in handling the equipment | Modify the size suitable for a single user | Is there any recommended dimension of the manual equipment |
| 6 | Reduce the bucket lifting effort | Implement overall load redistribution | Effort requirements should be suitable for a single worker |
| 7 | Beautification and visual appeal can be improved | Paint or coating can be implemented | No comment |
| 8 | An automatic lifting mechanism is required | Require a mobile power source and can increase the price and maneuvering effort required. | Target users are mostly in the outskirts and rural areas. Power requirements can hinder usage. |
| | Prototype Version 3 | | |
| 1 | Further reduction in the bucket lifting effort | Modify the link lengths | What should be the optimal link length? |
| 2 | An efficient waste disposal mechanism is required | Increase the flow rate of material | What should be the minimum slope for smooth dumping? |
| 3 | The hindered motion of equipment with a lowered bucket | The base plate of the bucket should not rub the ground | Bucket base plate can be made inclined for reduced contact |

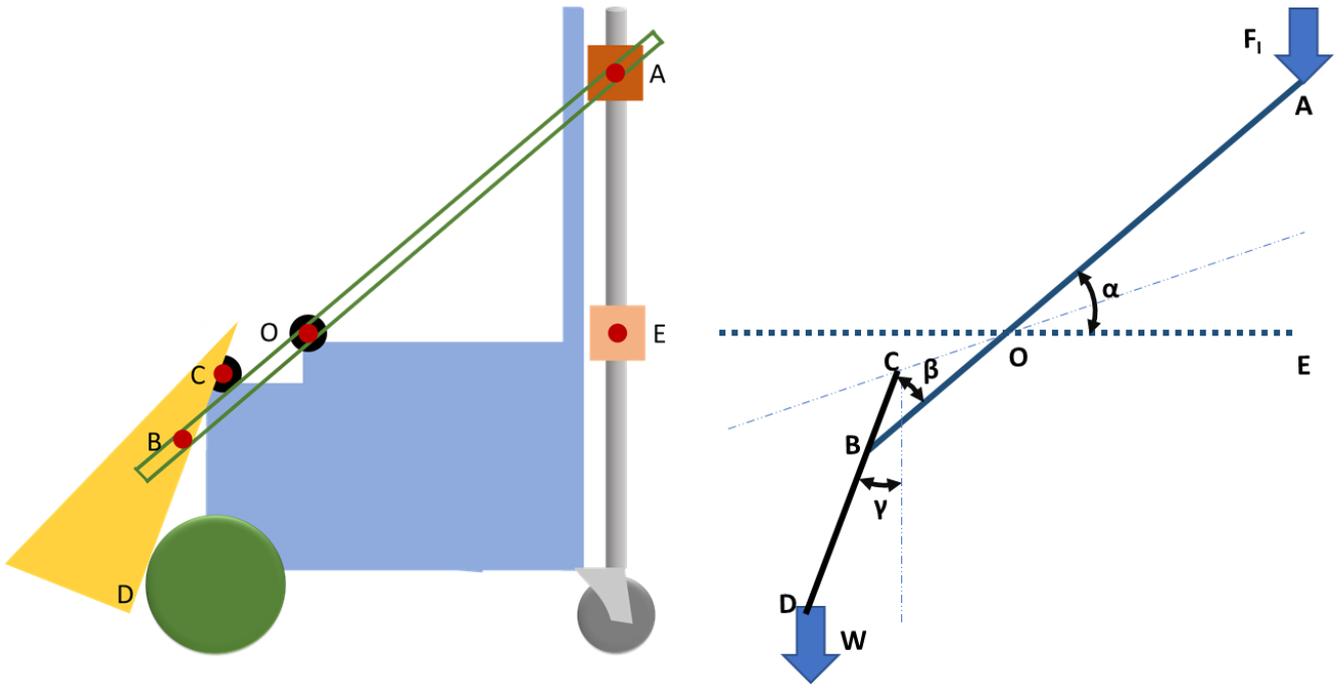

*Figure 4 Diagrammatic representation of the links and joints in the equipment*

$$f = \frac{F_I \times OE}{OB} \qquad Eq\ 1$$

$$W \times \sin\gamma = \frac{f \cos\beta \times CB}{(CB + BD)} \qquad Eq\ 2$$

Length of the link OB, variable due to slot joint (assuming OC=OB), is given by Eq 3.

$$OB = 2 \times CB \times \cos\beta \qquad Eq\ 3$$

The MA can be derived using Eq 1-3.

$$MA = \frac{OE}{2 \times (CB + BD) \times \sin\gamma} \qquad Eq\ 4$$

*2.3.1.2 Constraints*

The objective function (Eq 4) derived in the preceding section is constrained using anthropometric, feasibility, and usability requirements. The anthropometric constraint suggests that the lifting handle should be accessible to 95$^{th}$ percentile of the population. So, lifting handle height at A (Figure 5), considering the "comfortable vertical upward grasp reach from the floor" for 5$^{th}$ percentile Indian males as 1744 mm [29], should be less than 175 cm (Eq 5).

$$H_{\max.} = OE \tan \alpha + CB \left(1 + \sin \frac{\alpha}{2}\right) + BD \leq 175 \; cm \qquad Eq\;5$$

The feasibility constraint recommends the easy and smooth replacement of the existing cleaning equipment. So, the storage capacity of the new equipment should be at least equal to the commercial equipment with waste collection capacity of 60 liters [30]. The width of the equipment, considering the bi-deltoid width of the 95th percentile Indian male (482 mm) with clothes tolerance, should be more than 50 cm for easy access [29]. The minimum ground clearance and slider gap are assumed to be 15 and 5 cm respectively for smooth motion of the connecting parts resulting in Eq 6.

$$Volume_{Cont.} = Length\;(l) \times Width\;(w) \times Height\;(h) \geq 60000 \; cm^3$$

$$\left(OE + CB \cos \frac{\alpha}{2} - FE\right) \times 50 \times (CB + BD - GD) \geq 60000 \; cm^3 \qquad Eq\;6$$

The usability constraint governs the easy movement of the equipment. The equipment should be usable even on village roads, so the turning width (T) required (Figure 6) should not exceed 3 m [31]. The turning width should also incorporate the worker and its width must be added to the overall equipment length. The width of the worker, considering the chest thickness of the 95th percentile member of 25.4 cm [29], can be selected as 25 ensuing the final equipment length in Eq 7.

$$T = 2\sqrt{(w^2 + (l + 25)^2)} \leq 300 \; cm$$

$$OE + CB \cos \frac{\alpha}{2} \leq 115 \; cm \qquad Eq\;7$$

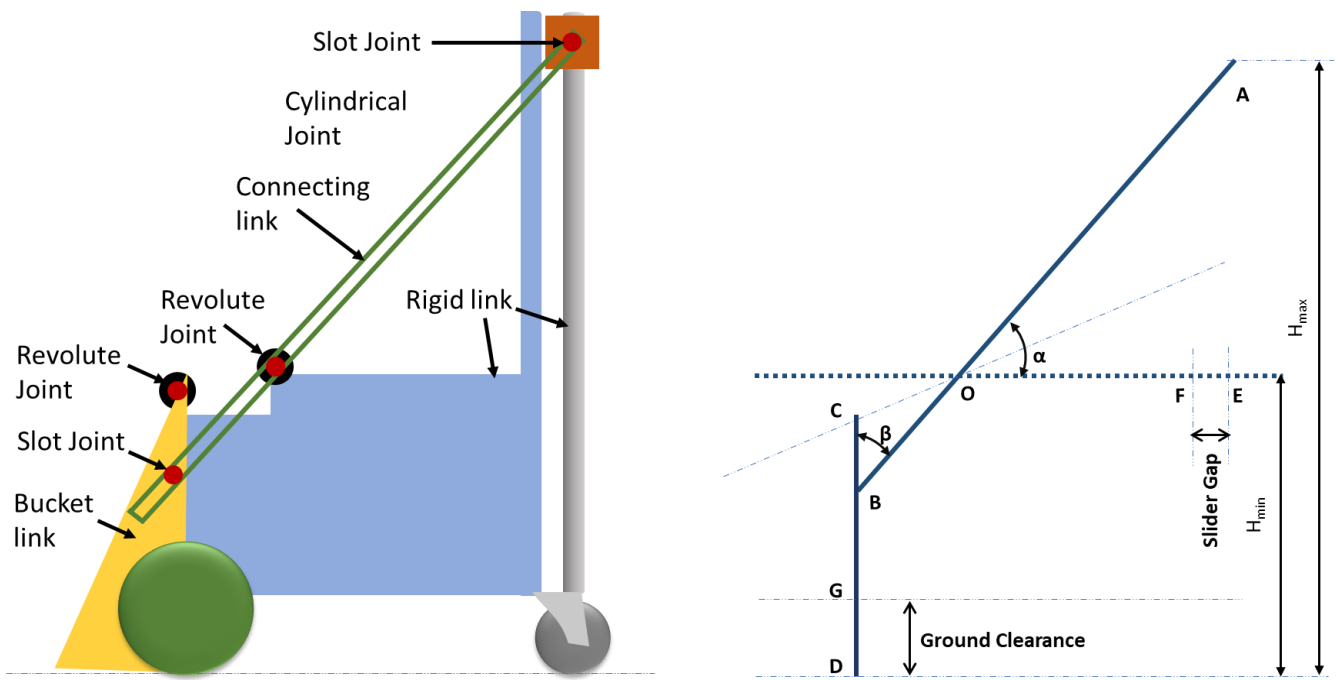

*Figure 5 Details of the lifting mechanism and its optimization parameters*

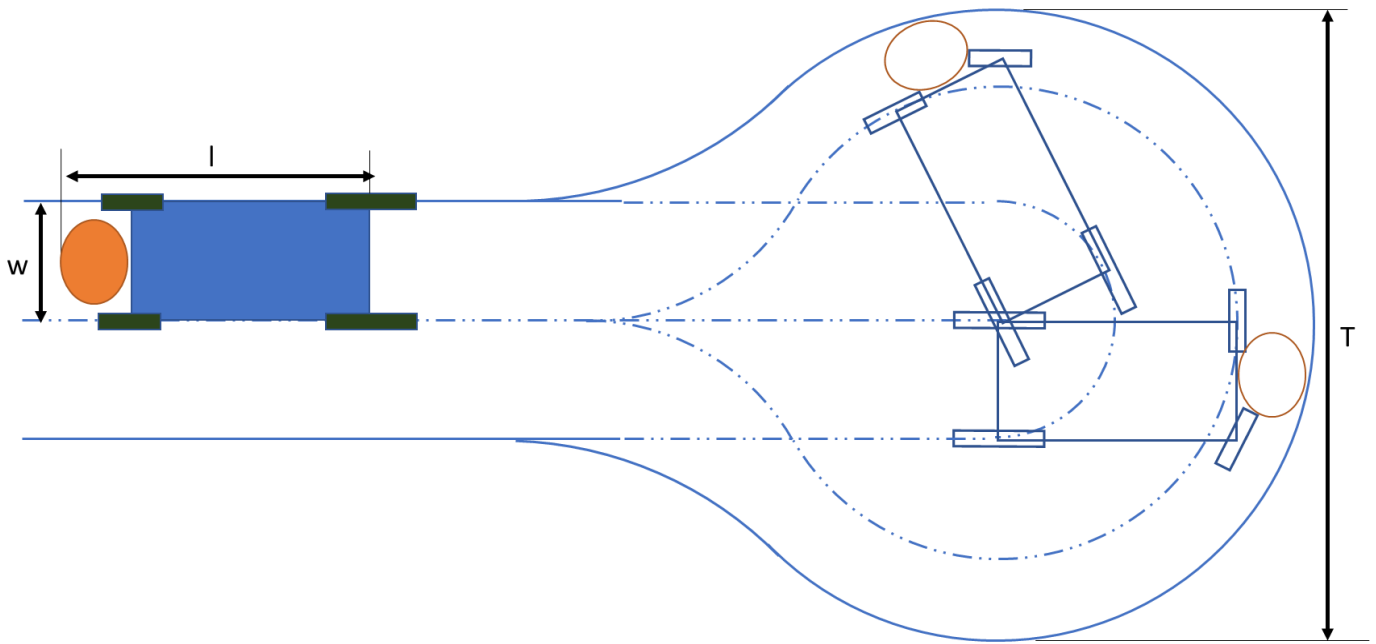

*Figure 6 Pictorial representation of the turning radius of the equipment*

## 2.3.1.3 Problem formulation

The optimization process aims to design the equipment with maximum efficiency with contemplation for workers' ergonomic threshold. The standard formulation of the optimization problem could be achieved using Eq 4-7. The value of γ varies from 0° (initial position) and goes up to 120° (considering the angle of repose for the material as 30°). So, the position of minimum MA with γ as 90° has been chosen for maximization in the objective function. The gradient-based nonlinear optimization function of MATLAB [32] is used for obtaining the solution as OE = 78 cm, CB =24 cm, and $BD$ =4 cm.

Obj fun

$$f = -\frac{OE}{2 \times (CB + BD)}$$

Sub to

$$OE \tan \alpha + CB \left(1 + \sin \frac{\alpha}{2}\right) + BD - 175 \leq 0$$

$$1200 - (OE + CB \cos \frac{\alpha}{2} - 5)(CB + BD - 15) \leq 0$$

$$OE + CB \cos \frac{\alpha}{2} - 115 \leq 0$$

$$-OE \leq 0$$

$$-CB \leq 0$$

$$-BD \leq 0$$

2.3.2  Handle position

*2.3.2.1 Lifting handle*

In the existing equipment, the lifting handle slides down from point A ($H_{max}$) to E ($H_{min}$) during the bucket lifting task with α decreasing from 60 to 0 degrees (Figure 7). The $H_{max}$ has been constrained in section 2.3.1.2 and the $H_{min}$ will be focused on in this section. $H_{min}$ position should be such that the 95th percentile of the population has comfortable access to it. To observe the variation of back compression force (BCF) with respect to the handle height, 3DSSPP software with 95th percentile male, 50th percentile mixed, and 5th percentile female anthropometry is used. As the lifting handle height goes below 80 cm, a sudden spike is evident from Figure 7. The stroke length ($H_{max} - H_{min}$) for the lifting handle is insufficient with 80 cm height. So, a tiered handle arrangement is used with 'between handle distance', equal to the forearm length of the 5th percentile member [29], as 20 cm.

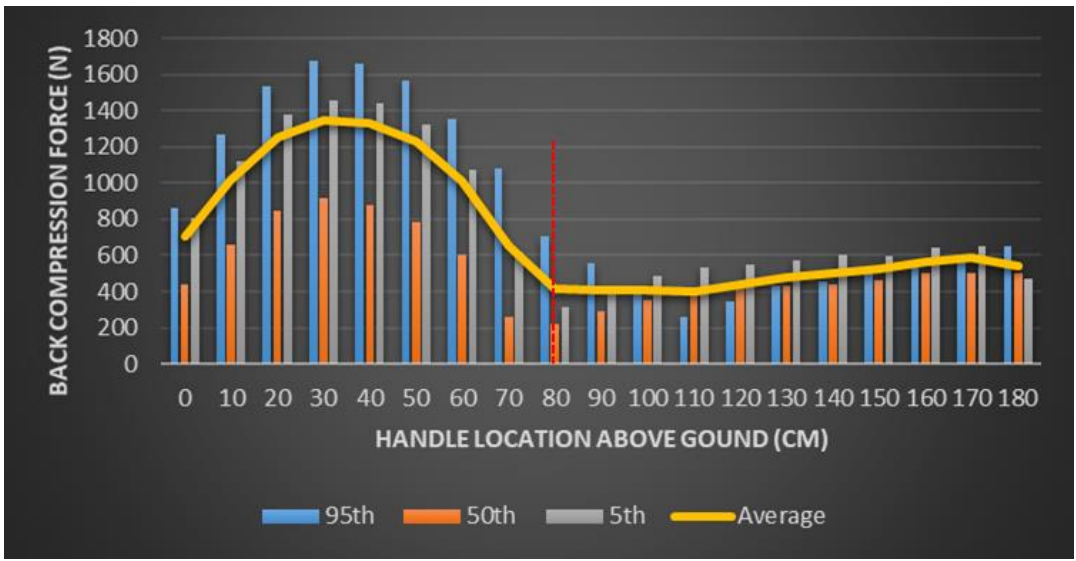

*Figure 7 Variation of back compression force with lifting handle height*

*2.3.2.2 Maneuvering handle*

The maneuvering handle should be designed such as to impose minimum stress on the body while performing the essential task. Two separate types of forces, namely push and pull, depends independently on the handle position. The DHM based study with acceptable maneuvering force for 5th, 50th, and 95th percentile members has been conducted. The study shows that minimum BCF for the pushing is achieved at higher (127.5) height while for pulling at lower (67.5) height (Figure 8). This indicates a tradeoff between the minimum push and pull handle heights, which has been resolved by accepting equally (67 percentile rank) good handle height of 92.5 cm for minimizing BCF.

2.3.3  Bucket dimension

Bucket height is already optimized in *section 2.3.1.3*. The width of the bucket will be equal to the width of the container with 5 cm bearing clearance, 45 cm. The length of the base plate of the bucket can be derived by analyzing its penetration depth into the material. Video analysis of the worker's scooping operation on the prototype using Kinovea software indicates the approach velocity to be 50 cm/sec. Taking 50 cm/s to be the initial velocity (u), final velocity (v) as 0 when the penetration stops, s as the distance covered by the base plate, the acceleration (a) can be calculated using Eq 8.

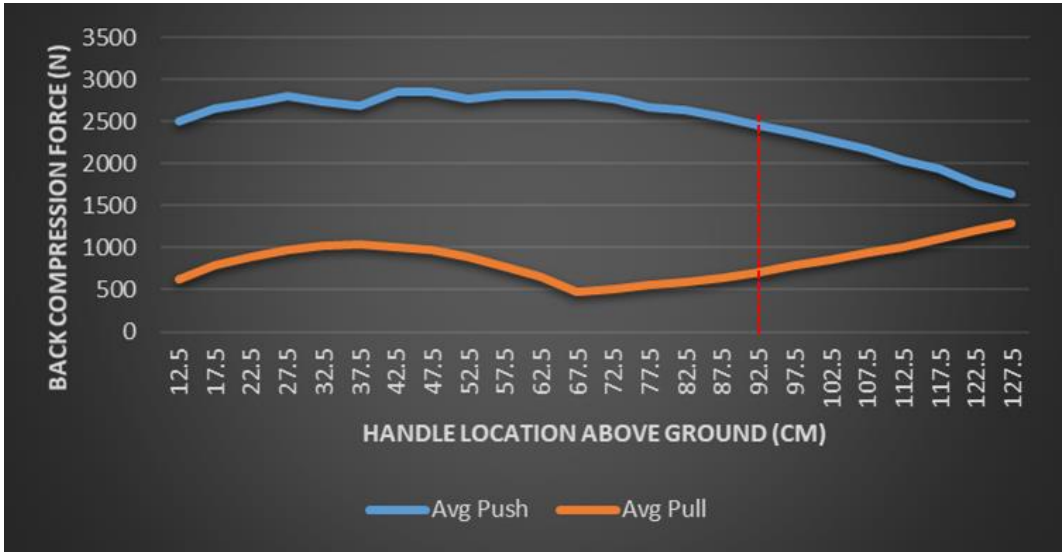

*Figure 8 Variation of the back-compression force with push-pull handle height*

$$v^2 - u^2 = 2as \qquad Eq\ 8$$

Also, the force required for scooping the material can be given by Eq 9.

$$F = \frac{m(u^2 - v^2)}{2s} \qquad Eq\ 9$$

The highest density material that the equipment could encounter is gravel. So, the calculations are according to gravel properties to incorporate the worst-case scenario. The resistance offered by the material is given by Eq 10.

$$F = \mu(\rho As)g \qquad Eq\ 10$$

where, μ is the coefficient of friction between two steel surfaces and gravel = 2 × 0.4 = 0.8 [33], ρ is the density of gravel= 2400 kg/m³ [34], A is the area of the bucket = $\frac{1}{2} \times CG \times 45$= 0.063 m². The resistance offered by the gravel should be equal to the effort applied. So, the effective length of the base plate of the bucket can be calculated using Eq 8-10

$$s = \sqrt{\frac{m(u^2 - v^2)}{2(\rho A)\mu g}}$$

*Eq 11*

Kerb mass of the equipment, m has been measured as 38 kg. The value of s is calculated using Eq 11 as 6.32, 8.98, 8.94, and 16.11 cm for gravel, mud, sand, and domestic waste respectively. The smallest length of the base plate can eventually be selected as 6 cm for its complete utilization.

### 2.3.4 Wheel selection

Scant information is available for the coefficient of rolling friction at the wheels and in bearings, the effect of the floor surface, and wheel diameter. So, an experimental study focusing on the effect of such factors on the maneuvering effort required has been conducted.

Three different 25 mm wide front wheels are tested: diameter 605, 655, and 695 mm and with each pair weighing 5.2, 6.3, and 11 kg respectively. This has been the diameter range available with the same shaft size, as the wheels are mounted on the same detachable axle. This axle has been used to restrict the increase in the height of the equipment while changing the front wheels. Two rear castor wheels with stopper are tested: diameter 155 and 240 mm and with each pair weighing 1.25 and 2.75 kg respectively. Changing castor wheels varied the height of the equipment so the height of the front wheels and maneuvering handle has been adjusted to accommodate the change. The testing has been done on two different materials generally used in Indian roads: asphalt and cement. Three different levels of material load are applied on the container: 0, 48.5 kg, 97 kg. Two direct effort types are also included: push and pull. Thrice repetitions of each factor level resulted in $3 \times 2 \times 2 \times 3 \times 2 \times 3 = 216$ experiments

The masses are measured using a digital hanging scale (Techmart, India) with full scale 200±0.02 kg. The manual push-pull efforts are measured using an electronic force gauge (Lutron electronic, Taiwan) with a full scale of 20±0.01 kg. A customized attachment at the rear end, to mimic the maneuvering handle, for applying effort, and a push-pull attachment in the front for interaction with the equipment is affixed on the force gauge. The equipment started from a stationary position for each measurement and the minimum force required to initiate the movement has been measured. The gauge has been attached at the midway along the equipment's width and the force is increased slowly and steadily without jerk.

The stepwise ($\alpha$=0.15) response surface method has been used for the regression analysis on Minitab 19 [35]. The alpha-to-enter significance level as 0.15 (higher than the usual 0.05) aided easy entry of predictors into the model, while alpha-to-remove as 0.15 restricts their easy removal. So, setting the higher alpha value helps to analyze the impact of predictor variables in the regression model. The large $R^2$ value (95.19) suggests that the variation in the effort is reasonably predictable from the factors considered.

The p-values in the ANOVA [36] analysis (Table 2) suggest that the rear wheel, material weight, and floor type significantly influence the effort value. The full effect of interaction between the factors on the minimum effort required for initiating the movement in the equipment is shown in Figure 9. The optimal desirability for the low effort includes smaller diameters for both front and rear wheels, less material load, cemented road, and pulling tasks (Figure 10).

*Table 2 The p-values for the factors*

| Factors | p-value |
|---|---|
| Rear-wheel (*RW*) | 0.000* |
| Front-wheel (*FW*) | 0.088 |
| Material weight (*MW*) | 0.000* |
| Floor-type (*FT*) | 0.000* |
| Effort type (*ET*) | 0.073 |
| $FT \times FT$ | 0.033* |
| $MW \times MW$ | 0.026* |
| $RW \times MW$ | 0.000* |
| $FW \times ET$ | 0.015* |
| $MW \times FT$ | 0.003* |
| $FT \times ET$ | 0.009* |

*Significant at p<0.05

2.3.5   Dumping mechanism

The final feature that has been optimized is the disposal system. The slope in the base plate and side opening door didn't amount to the smooth material flow in prototype version 3. The study to measure the angle of repose in the final bucket angle ($\gamma$) as 30°, has been utilized for determining the slope of the disposal door. Considering the ground clearance of only 15 cm, a double door hopper-type system is installed at the base. The self-locking latch enabled easy and quick material disposal.

# 3   Results

The results of each of the steps of the methodology are already discussed in the previous sections. In this section, the focus will be on the evaluation of the ergonomically modified equipment. Also, a comparative overview of the different versions of the equipment, that are evolved after each iteration using the methodology, will be presented.

The CAD model of the optimized design has been prepared using SolidWorks 2020 [37] (Figure 11) and the mechanical and material feasibility of the equipment is done through ANSYS 2019 [38]. After the initial check, prototype version 4 has been manufactured for ergonomic assessment. The minimum mechanical advantage is significantly improved to 3.75 (241%) in the latest version when compared with the previous versions, i.e., version 1: 0.3; version 2: 0.5 (67%); version 3: 1.1 (120%), owing to the smaller bucket size and improved load distribution.

The posture attained while manipulating the equipment's mechanism has been studied through the ergonomic posture assessment tool (Table 3). (i) REBA analysis suggested that the initial posture score is improved from 3 to 2 in the 4th version of the prototype. The REBA score identified the final postures in the initial versions 1-2 as the poorest postures with high MSDs risk. The overall REBA score for the final manipulation task of version 4 is well under 3, which is a 6-points reduction compared to version 1.

The biomechanical study using 3DSSPP software substantiated the feasibility of each of the design variants discussed in the brainstorming session. Different versions of the modifications were initially analyzed, and the best version has been implemented for the prototype manufacturing. The comparative analysis of the prototype version 1-4 are presented in Table 3. The back compressive force (BCF) in the user's spine for the final posture is minimum in version 4. The Strength Percent Capable (%SPC) gives the percentage of the population with enough strength to sustain the posture and not be fatigued over the work shift's length. The %SPC analysis suggested that only 38% of the population will be able to accomplish the bucket lifting task in version 1 and will face body balance issues in version 2. The analysis of versions 3 and 4, indicated a steep rise in %SPC to 98 and 99% respectively.

The 3DSSPP software can also be used to determine the percentage of maximum voluntary contraction of the muscles (%MVC) associated with the body's joint under study. The study suggested that version 1 had a staggering 140 %MVC for the wrist joint during the final manipulation posture, while %MVC for version 2 is not identifiable due to the imbalance in the posture during foot operations. The %MVC for all the joints for equipment manipulation in version 3 and 4's are well under 34% and 27% respectively.

## 4 Discussion

In this study, a new methodology for integrating ergonomics in design is presented. The empirical formulation has been derived from the principles of human-centered design and exemplified as an easy and practical approach for the development of manual equipment. The methodology mimics the ideal practice of maximizing productivity while maintaining the ergonomic risks below the threshold levels. This approach presents the novelty in its flexibility of the method utilized in performing each of the steps. The complexity of the approach often restricts its application, while a single rigid approach acts as a bottleneck. Most of the previous studies [2,39] often utilize a setup or software which upsurges the dependency of the methodology on that initial investment or skill. The steps discussed in the present study could be performed, if time and money permits, or could be achieved using alternatives suggested. This factor enhances the utilization quotient of the methodology multifold.

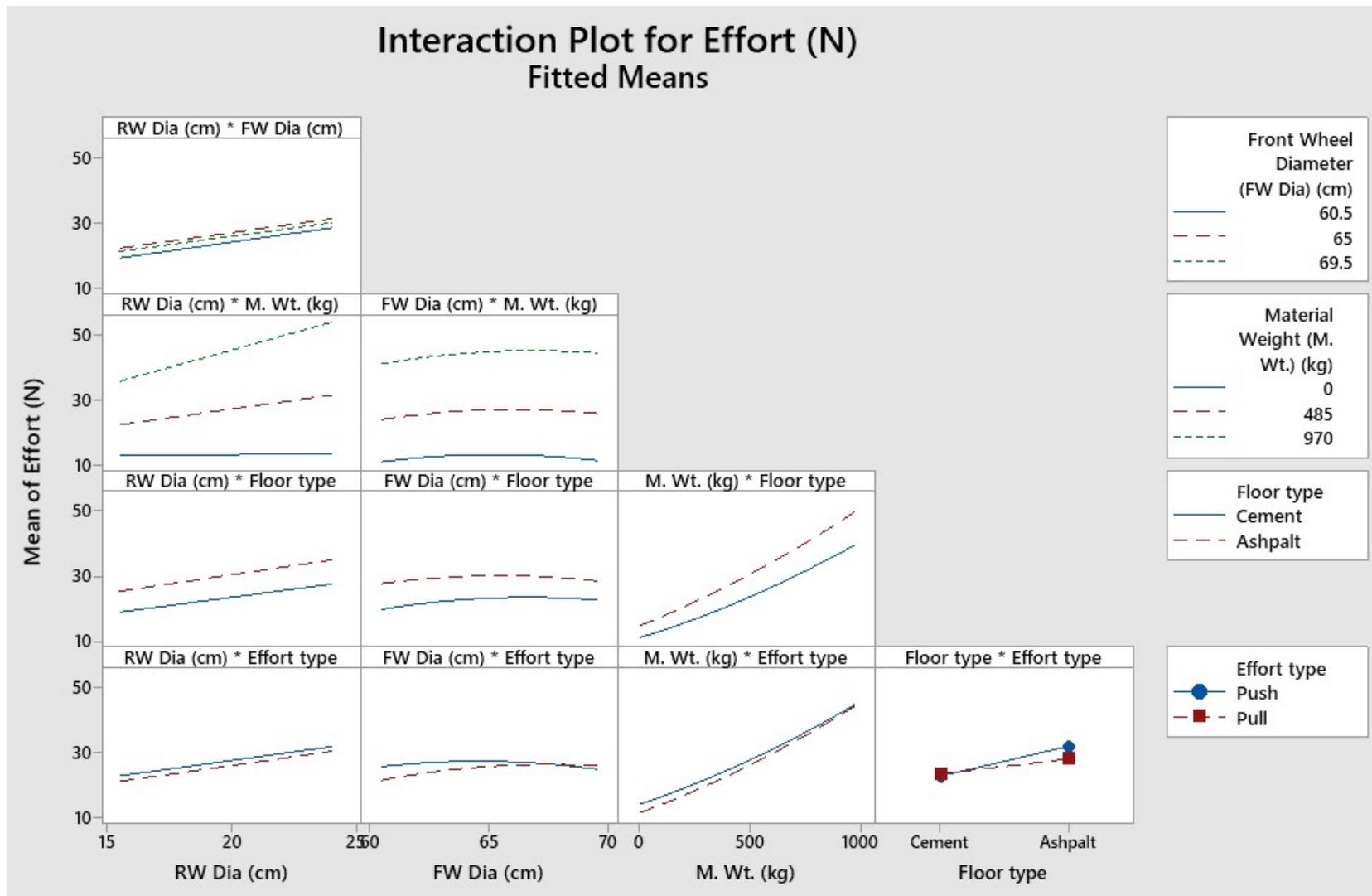

*Figure 9 The effect of interaction between factors on the effort required for maneuvering the equipment*

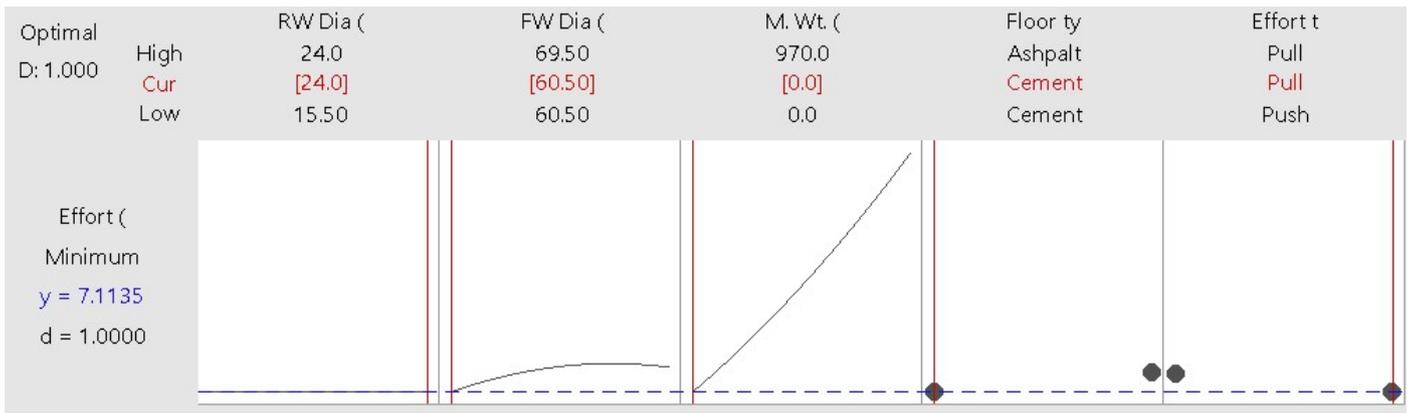

*Figure 10 Selection of the optimal value of the factors for minimum effort*

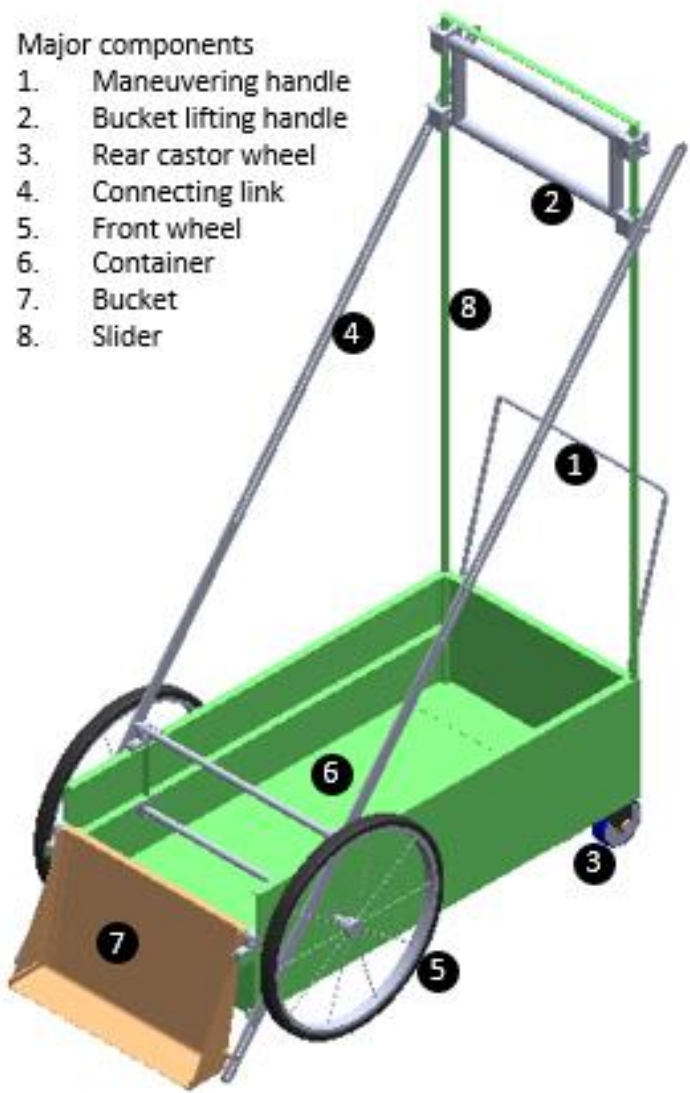

*Figure 11 CAD model of the cleaning equipment*

The other advantage of this methodology is the continuous improvement loop. Flowchart discussed in the previous studies often closes by an end that gives a false perception of achieving perfection [7,40]. Efficiency and human well-being are desirable factors and can never achieve perfection. Four versions of the cleaning equipment are discussed in this paper and undoubtedly each version is an upgrade over its previous self, still, there will be improvements in the latest version as well in the future.

The study here mainly focuses on the optimization of the design parameters but the motive of introducing the methodology is to give insight into the whole development process. The researchers often shift to the design optimization step directly which might cover only half of the picture [41]. The search for the robust parameters required for the optimization begins with the survey. The involvement of the user is unswerving and leads to the identification of the principal issue with the system and tools [42]. Such a survey also gives an overview of the anthropometry of the user which is unquestionably the key parameter in integrating ergonomics in design [43]. The survey for equipment is a must while developing equipment from scratch, and can also be assistive in introducing techniques and technologies of totally unrelated fields. The process of continuous improvement often heads towards a dead-end and in such field-specific cases, interdisciplinary studies could act as a boon by introducing a new dimension to the search [44]. It is this area evidently, where optimization fails but the presented methodology leads to a novel device.

Another major step in the present study is the brainstorming sessions as discussed in the methodology. Brainstorming sessions are the source of innovative or outlandish ideas that are eventually molded into novel designs [45]. The involvement of the end-user throughout the process reduces the chances of rejection of the new interventions. The existing designs could also be adequately intervened using a user perspective [46], which forms the basis of ergonomic intervention and inclusion of human factors

*Table 3 Summary of posture based ergonomic assessment and biomechanical analysis*

| Prototype version 1 | | | Prototype version 2 | | | Prototype version 3 | | | Prototype version 4 | | |
|---|---|---|---|---|---|---|---|---|---|---|---|
| | REBA | 3 | | REBA | 3 | | REBA | 1 | | REBA | 2 |
| | BCF | 685 N | | BCF | 621 N | | BCF | 642 N | | BCF | 700 N |
| 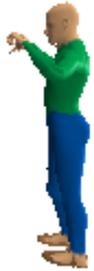 | *Part* | *SPC MVC* | 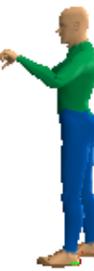 | *Part* | *SPC MVC* | 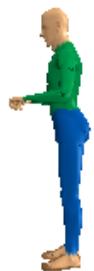 | *Part* | *SPC MVC* | 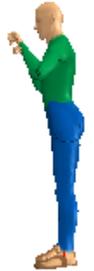 | *Part* | *SPC MVC* |
| | Wrist | 100  3 | | Wrist | 100  4 | | Wrist | 100  5 | | Wrist | 100  3 |
| | Elbow | 100  0 | | Elbow | 100  3 | | Elbow | 100  5 | | Elbow | 100  2 |
| | Shoulder | 100  20 | | Shoulder | 100  20 | | Shoulder | 100  11 | | Shoulder | 100  15 |
| Initial posture | Torso | 100  16 | Initial posture | Torso | 100  13 | Initial posture | Torso | 100  12 | Initial posture | Torso | 99  15 |
| | Hip | 99  11 | | Hip | 98  27 | | Hip | 99  8 | | Hip | 100  13 |
| | Knee | 98  29 | | Knee | 100  30 | | Knee | 100  9 | | Knee | 100  14 |
| | Ankle | 95  60 | | Ankle | 99  23 | | Ankle | 100  10 | | Ankle | 100  13 |
| | REBA | 9 | | REBA | L-7  R-8 | | REBA | L-2  R-1 | | REBA | L-2  R-1 |
| | BCF | 723 N | | BCF | 1595 N | | BCF | 616 N | | BCF | 583 N |
| 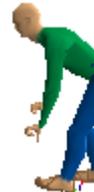 | *Part* | *SPC MVC* | 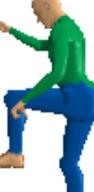 | *Part* | *SPC MVC* | 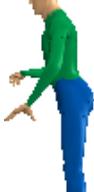 | *Part* | *SPC MVC* | 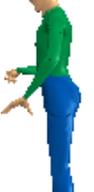 | *Part* | *SPC MVC* |
| Final posture | Wrist | 38  140 | Switching posture | Wrist | 100  5 | Switching posture | Wrist | 99  34 | Switching posture | Wrist | 100  1 |
| | Elbow | 100  45 | | Elbow | 100  4 | | Elbow | 100  16 | | Elbow | 100  3 |
| | Shoulder | 100  11 | | Shoulder | 100  22 | | Shoulder | 100  9 | | Shoulder | 100  8 |

| Prototype version 1 | | | | Prototype version 2 | | | | Prototype version 3 | | | | Prototype version 4 | | |
|---|---|---|---|---|---|---|---|---|---|---|---|---|---|---|
| | Torso | 100 | 16 | | Torso | 99 | 30 | | Torso | 100 | 12 | | Torso | 100 | 10 |
| | Hip | 96 | 43 | | Hip | NA | NA | | Hip | 99 | 9 | | Hip | 99 | 6 |
| | Knee | 99 | 31 | | Knee | NA | NA | | Knee | 100 | 10 | | Knee | 100 | 8 |
| | Ankle | 99 | 23 | | Ankle | NA | NA | | Ankle | 100 | 11 | | Ankle | 100 | 8 |
| | | | | | REBA | L-9 | R-8 | | REBA | 4 | | | REBA | 3 | |
| | | | | | BCF | 1533 N | | | BCF | 1110 N | | | BCF | 895 N | |
| | | | | | *Part* | *SPC* | *MVC* | | *Part* | *SPC* | *MVC* | | *Part* | *SPC* | *MVC* |
| | | | | | Wrist | 100 | 4 | | Wrist | 100 | 18 | | Wrist | 100 | 1 |
| | | | | | Elbow | 100 | 4 | | Elbow | 100 | 11 | | Elbow | 100 | 2 |
| | | | | | Shoulder | 100 | 24 | | Shoulder | 100 | 6 | | Shoulder | 100 | 2 |
| | | | | Final | Torso | 99 | 35 | Final | Torso | 99 | 27 | Final | Torso | 100 | 21 |
| | | | | posture | Hip | NA | NA | posture | Hip | 98 | 22 | posture | Hip | 99 | 15 |
| | | | | | Knee | NA | NA | | Knee | 100 | 4 | | Knee | 100 | 16 |
| | | | | | Ankle | NA | NA | | Ankle | 99 | 21 | | Ankle | 100 | 16 |

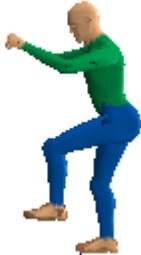
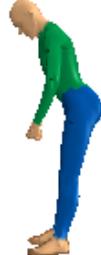
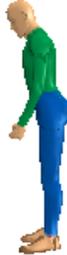

REBA: Rapid Entire Body Assessment; BCF: Back Compressive Force; L & R: Left and right; %SPC: Percentage of Strength Percent Capable; %MVC: Percentage of Maximum Voluntary Contraction in design. In this study during the development phase, many such ideas were primed by cleaning workers, which after DHM based analysis, are found to be quite effective and cost-saving as well.

Despite the advantages, this study has its limitations. Firstly, the proposed methodology tries to include the robustness as well as the flexibility to provide a generalized tool, which may appear slightly complex to novice users. Secondly, the increase in the number of steps in the methodology could give an impression of increased time consumption. Thirdly, the methodology only indicates the objectives and the parameters for optimization and not precisely the technique most suitable for it. In the paper, mechanical advantage has been selected as the criteria for mechanical optimization which does not exactly capture the effects of friction and moment of inertia. In the future, multibody dynamics simulations may be used for evaluating the objective function and thereby improving the equipment design. The above-suggested limitations may be ignored as the outcome of the study is highly productive, beneficial, and user-friendly.

## 5 Conclusion

The eight-step methodology has been provided based on the experience acquired during the development of the equipment. The experimental design helped in successfully introducing the enhanced cleaning equipment among workers. The authors strongly believe that the methodology could act as a guide in the development to deployment stages of manual or semi-automatic equipment. The direct and simple parametrization in the optimization technique can promote researchers or budding ergonomists to include optimization tools while developing low-cost equipment as well.

## 6 Relevance to industry

The methodology could help reduce space and time which will eventually lead to affordable tools and equipment for conventional and modern small to medium scale industries. Such affordable equipment can improve the workplace condition of the workers multifold, which is going to be the future demand of the industries.